\documentclass[preprints,article,accept,pdftex,moreauthors]{mdpi}

\firstpage{1} 
\pubvolume{1}
\issuenum{1}
\articlenumber{0}
\pubyear{2022}
\copyrightyear{2022}
\datereceived{} 
\dateaccepted{} 
\datepublished{} 
\hreflink{https://doi.org/} 
\pdfoutput=1

\Title{High energy and high power primary Li-CF$_{x}$ batteries enabled by
the combined effects of the binder and the electrolyte}

\TitleCitation{Title}

\Author{Haobin Huo $^{1}$, Leon L. Shaw $^{1}$ and K{\'a}roly N{\'e}meth $^{2,}$*}

\AuthorNames{Haobin Huo, Leon L. Shaw and Karoly Nemeth}

\AuthorCitation{Huo, H.; Shaw, L. L.; Nemeth, K.}

\address{%
$^{1}$ \quad Mechanical, Materials, and Aerospace Engineering Department;
hhuo2@hawk.iit.edu, lshaw2@iit.edu \\
$^{2}$ \quad Physics Department, Illinois Institute of Technology; knemeth@iit.edu}

\corres{Correspondence: knemeth@iit.edu}

\abstract{
Several effective methods have been developed recently to demonstrate
simultaneous high energy and high power density in Li - carbon fluoride
(CF$_{x}$) batteries. These methods can achieve as high as ~1000 Wh/kg energy
density at ~60-70 kW/kg power density (40-50 C rate) in coin cells and
~750 Wh/kg energy density at 12.5 kW/kg power density (20 C rate) in
pouch cells. This performance is made possible by ingenious
nano-architecture design, controlled porosity, boron doping and 
electrolyte additives. In the present study, we show that
a similarly great performance, 931 Wh/kg energy density at 59 kW/kg
power density, can be achieved by using a polyacrylonitrile
binder and a LiBF$_{4}$ electrolyte in Li - graphite fluoride coin cells. 
We also demonstrate that the observed effect is the result of
the right combination of the binder and the electrolyte.
We propose that the mechanistic origin of the observed phenomena
is an electro-catalytic effect by the polyacrylonitrile binder.
While our proposed method has a competitive performance, 
it also offers a simple implementation 
and a scalable production of high energy and high power 
primary Li-CF$_{x}$ cells.
}


\begin{document}


\section{Introduction}

Functionalized two-dimensional (2D) materials as cathode active
species have a demonstrated ability to realize simultaneously high 
energy and high power in primary and secondary batteries. These
features are highly desirable for many applications ranging
from electric cars to electric aircraft, space exploration, 
pulsed power sources and medical devices 
\cite{trahey2020energy,yang2021challenges,bills2020performance,krause2021performance,krause2018high,ndzebethigh2018,greatbatch1996lithium}.
Graphite fluoride (CF$_{x}$, 0<x<~1.3) is the oldest known example of 
functionalized 2D materials as cathode active species.
The structure of fully fluorinated graphite is known since 1947
and consists of stacked fluorinated layers of graphene 
\cite{ruff1934reaktionsprodukte,rudorff1947konstitution}.
The Li-CF$_{x}$ primary battery was commercialized in 1970 
\cite{watanabe1970primary,watanabe1972high,fukuda1981active}.
It offers many advantages such as high energy and high power
density, excellent shelf life, applicability in a wide temperature range
($-$60 to $+$60 $^{o}$C) and a relatively easy to source and
economic composition
\cite{sharma2021fluorinated,ahmad2021preparation,wang2022composite,groult2018use,zhang2015progress}.
It has a very high theoretical specific energy
of 2180 Wh/kg at a capacity of 864 mAh/g 
when graphite is fully fluorinated (x$\approx$1)
and an open circuit voltage (OCV) of 3.2-3.3 V 
\cite{watanabe1970primary}. It is a primary battery with only
a limited degree of rechargeability 
\cite{chen2021fluorinated,liu2014rechargeable}.
The cell reaction on discharge is the following:
\begin{equation}
x Li^{+} + x e^{-} + CF_{x} \rightarrow x LiF + C.
\end{equation}
If the discharge product carbon would be graphite, an OCV 
of 4.57 V should be observed on the 
basis of thermodynamical calculations
\cite{whittingham1975mechanism,watanabe1987discharge}.
The much lower observed OCV is a consequence of the formation of
a sandwich structure of LiF and graphene in the discharge product
\cite{whittingham1975mechanism,watanabe1982solvents,watanabe1987discharge}. 

The study of solvent effects led to the recognition that the solvated
Li$^{+}$ ion intercalates between the stacked CF$_{x}$
monolayers during discharge and the solid LiF discharge product forms 
only after the collapse of the solvent shell of Li$^{+}$ 
between the CF$_{x}$
layers \cite{watanabe1982solvents,watanabe1987discharge}. The
higher the solvation energy of Li$^{+}$ in a given solvent the lower
the discharge voltage \cite{watanabe1982solvents}.

The intercalation of solvated Li$^{+}$ ions was also seen in 
graphene oxide (GO) cathodes. In order to achieve high
power density in Li/Na-GO batteries, the GO interlayer distance must
be expanded, which is typically achieved via a thermal treatment
of GO before building the cathode
\cite{jang2011graphene,liu2014lithium,kim2014all,kim2014novel,kornilov2022li}.
The layer distance in fully fluorinated CF$_{x}$ is about 6-9 {\AA}
\cite{rudorff1947konstitution}
and the interaction between the layers is weak enough to allow for the
penetration of some solvated Li$^{+}$ ions.
The discharged CF$_{x}$ (which includes the intercalated
CF$_{x}$ and its discharge products) 
typically forms a shell around a core of 
CF$_{x}$ particles and has a major impact on the
overall performance of the Li-CF$_{x}$ cells \cite{zhang2009electrochemical}.
The time evolution of the structure and composition of the
cathode during discharge has been studied recently in detail at a slow 
discharge rate \cite{sayahpour2022revisiting}. As opposed to
expectations, the graphene layers transform into a hard carbon
structure during discharge with a much reduced sp$^{2}$ carbon
content. The optimization of the structure and composition of the 
domain of the 
discharged CF$_{x}$ is the key to improving the electrochemical
performance of Li-CF$_{x}$ cells.

The specific power of Li-CF$_{x}$ cells used to be small
historically. 
In 2007, commercial Li-CF$_{x}$ batteries were reported to have
a break-down of specific energy at a specific power of 
$\approx$1.6 kW/kg which was cured by the introduction of partially
fluorinated CF$_{x}$ ($\approx$0.3<x<$\approx$0.8) cathodes. 
Partially fluorinated CF$_{x}$ could realize 
much higher specific power values and a break-down of specific energy at
10 kW/kg \cite{yazami2007fluorinated,lam2006physical}.
Research on high power Li-CF$_{x}$ batteries intensified in the
past decade. Various approaches have been developed which are
capable to deliver a specific energy of 800-1000 Wh/kg at power
densities of 20-70 kW/kg (15-50 C rates). These approaches utilize 
nano-architecture
design, such as fluorinated graphene microspheres \cite{luo2021ultrafast50C},
increased porosity and edge (instead of in-plane)
functionalization \cite{dai2014surface30C,peng2021fluorinated20C,jiang2021electrochemical20C,wangfluorination20C} 
and amorphization of the discharge product LiF \cite{li2021gaseous15C}.
Recently, the amorphization of LiF was achieved by the addition of
BF$_{3}$ gas in the electrolyte at a 0.01 M concentration
and was confirmed by the
lack of LiF pattern in the X-ray diffraction pattern of the
discharged CF$_{x}$ cathode \cite{li2021gaseous15C}. 
Amorphization of LiF was also observed
when a solid electrolyte, Li$_{3}$PS$_{4}$ was used 
\cite{rangasamy2014pushing}.

Another method of amorphization of the discharge product LiF 
was proposed by Jones and Hossain
about a decade ago as a way to reduce the relatively high 
heat production during
discharge \cite{jones2011polymer}. These inventors proposed to use
polymeric binders that serve a dual purpose: (1) they mechanically bind
components of the cathode together and (2) act as complexation $/$
amorphization agents of LiF. Several such polymers were proposed as a
replacement of the more traditional poly( vinylidene difluoride) (PVDF)
and teflon (PTFE),
among them polyacrylonitrile (PAN) derivatives (primarily
complexing the Li$^{+}$ ions) and boronates (primarily
complexing F$^{-}$ ions). These complexation phenomena are
typically based on Lewis acid-base reactions.
The gaseous BF$_{3}$ electrolyte 
additive mentioned above is a simple example of the formation of
a Lewis adduct of LiF with BF$_{3}$
as BF$_{3}$ is well known to easily dissolve
LiF in the form of LiBF$_{4}$. Unfortunately, the theoretical
proposal of Jones and Hossain has not been experimentally
demonstrated to date, to the best of our knowledge. 

We have been working on the development of high energy and high power 
batteries based on functionalized 2D materials for about a decade
\cite{nemeth2014materials,zhang2016experimental,nemeth2018simultaneous,nemeth2021radical,nemeth2020-US10693137,nemeth2022radicalAnion-US11453596B2}.
We recently proposed the use of adducts of BF$_{3}$ with graphene oxide
(GO) and oxidized hexagonal boron nitride
(hBN) as cathode active species and solid electrolytes
\cite{nemeth2018simultaneous,nemeth2021radical,nemeth2020-US10693137,nemeth2022radicalAnion-US11453596B2}.

The above mentioned successful application of the BF$_{3}$ gas additive in 
Li-CF$_{x}$ cells \cite{li2021gaseous15C} has called our attention to the 
role of BF$_{3}$ and its
adducts in high power Li-CF$_{x}$ cells. However, our initial attempt to use
BF$_{3}$-etherate as a safer alternative to BF$_{3}$ gas in
Li-CF$_{x}$ cells to achieve high power density has failed. This has motivated us to
investigate alternatives. We have explored the use of LiOX$\cdot$BF$_{3}$ (X
= Li, H) to substitute -F to -OBF$_{3}$ groups in CF$_{x}$
for a greater interlayer spacing. As the addition of LiOX$\cdot$BF$_{3}$
led to an alkaline cathode, we have thus also investigated a PAN binder 
instead of the traditional PVDF because the latter 
is not stable in alkaline
environment \cite{marshall2021solubility}. This study has
resulted in the discovery of the
beneficial effects of a PAN binder on the power density 
of Li-CF$_{x}$ cells. Furthermore, we have discovered that such a 
binder effect occurs only in selected electrolytes. While the mechanistic 
origin of the combined binder and electrolyte effects are not clear yet, we
propose that it is likely related to the electro-catalytic activity of PAN.
The details of our findings are described below.

\section{Materials and Methods}
\begin{figure}[H]
\includegraphics[width=10.5 cm]{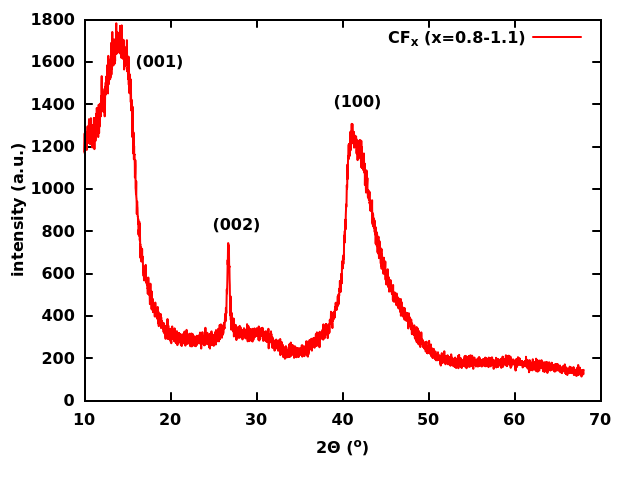}
\caption{
XRD (Cu K$\alpha$) pattern of the CF$_{x}$ used in the present
study. The presence of some unfluorinated graphite is
indicated by the (002) reflection.
\label{XRDCFX}
}
\end{figure}

Graphite fluoride (CF$_{x}$) was purchased from ACS Material (product
number GT1FS012) with the F/C ratio being 0.8-1.1. 
The X-ray diffraction (XRD) pattern (at Cu K$\alpha$ X-ray radiation) 
of this CF$_{x}$ is shown in Fig
\ref{XRDCFX}. The (002) reflection at 26.7 deg indicates the presence of
some unfluorinated graphite species.          

Lithium tetrafluoroborate (LiBF$_{4}$), lithium perchlorate (LiClO$_{4}$), 
lithium hexafluorophosphate (LiPF$_{6}$), lithium bis(oxalato)borate
(LiB(C$_{2}$O$_{4}$)$_{2}$, LiBOB), 
polyacrylonitrile (PAN, average molecular weight 150000),    
carbon black (CB), ethylene carbonate (EC), dimethyl carbonate
(DMC), propylene carbonate (PC), 1,2 dimethoxy ethane (DME),
dimethyl sulfoxide (DMSO), 1,3-dioxalane (DOL),
N,N-dimethylformamide (DMF) and N-Methyl-2-pyrrolidone (NMP) 
were purchased from Millipore-Sigma. 
Poly( vinylidene difluoride) (PVDF) was purchased from Alfa Aesar.
Conductive carbon black (CB, TIMCAL Graphite \& Carbon Super P) was
purchased from MTI Corporation.
the as purchased LiBOB was dried in vacuum at 150 $^{\rm{o}}$C for 8 h before use for better
solubility in electrolyte solvents \cite{zor2021guide}.

Cathodes were cast on an Al-foil from a slurry using variable
spreading gaps (100 and 250 ${\mu}m$). The slurries were made using
a mixture of CF$_{x}$, binder and carbon black at a mass ratio
of 8:1:1, respectively, unless otherwise noted. 
The CF$_{x}$ was ultrasonicated for 4 hours in ethanol (EtOH) following
similar ultrasonication processes of the 
literature \cite{zhang2013two,zeng2018dynamic}  
for exfoliation and easier homogenization of the slurry.
A Branson Ultrasonics CPX952516R device was used that allows for
mild ultrasonication (40 kHz).
The binders were added in the
form of a 4 w\% PAN/DMF or PVDF/NMP solution to the dry solid
CF$_{x}$ and CB and mixed thoroughly in a laboratory mixer
(Thinky ARM-310). The cathodes were dried in a vacuum oven at
120 $^{o}$C for 8h. A more detailed account on our laboratory
procedures is available in Ref \cite{huo2022high}.

The LiOX$\cdot$BF$_{3}$ additive \cite{tatagari2021functionalized}
was synthesized in the presence of CF$_{x}$ by 
mixing CF$_{x}$ with Li$_{2}$CO$_{3}$ in a 20:1 molar ratio and
ultrasonicating the mixture in F$_{3}$B$\cdot$OEt$_{2}$ for 4 h.
The product was washed with dichloromethane and filtered on a
nanoporous filter.
Products were dried in vacuum after ultrasonication.

Cathode disks of 0.65 cm radius were punched out of the cathode
sheets and placed into CR2032 coin cell cases. 
Porous polypropylene disks (Celgard 2500, 25 $\mu$m thickness) 
were used as separator and Li-foil disks as anodes. For electrolyte, 
a 1 M solution of LiBF$_{4}$, LiClO$_{4}$, LiPF$_{6}$ or
LiBOB was used in a 1:1:1 volumetric mixture of PC, DME and DOL
solvents following a similar electrolyte in Ref \cite{li2022multi}. 
The advantages of using DOL as an
electrolyte (co)solvent are discussed in Ref
\cite{zhao2020designing}.
Typically, a 60 ${\mu}L$ electrolyte was filled
in a coin cell. The cells were hermetically sealed using a
crimping machine.

Galvanostatic cycling of coin cell batteries was carried out
using a Neware battery tester (maximum voltage 5V, maximum
current 50 mA). The voltage limits were 1.5 and 4.6 V, unless
otherwise noted.
Electrochemical Impedance Spectra (EIS) were measured
using a PARSTAT 4000 instrument between 0.1 Hz and 100 kHz 
at a 10 mV amplitude.
XRD was carried out using a Bruker D2 Phaser device at Cu
K$\alpha$ X-ray radiation (1.5406 {\AA} wavelength). 
Additionally, synchrotron XRD (at 0.459063 {\AA} wavelength) was
also carried out at the Advanced Photon Source at Argonne
National Laboratory.

\section{Results and Discussion}

We tested a series of Li-CF$_{x}$ cells to understand
the binder and electrolyte dependence of the power density.
The compositions of these cells are listed in Table \ref{HXXX}.

\begin{table}[H] 
\caption{Cathode composition in different groups of cells.
Blends of solvents were obtained by mixing 
equal volumetric amounts of the components. CF$_{x}$ was
ultrasonicated for 4 h before use except in the case of
H304 (0 h) and H101 (3 h). 
The active material loading was 1.0-1.5 mg/cm$^{2}$ for 100
${\mu}$m thick cathodes and 3.0-4.6 mg/cm$^{2}$ for 250 ${\mu}$m
ones. 
The concentration of the electrolytes was always 1 M.
The molar ratio of CF$_{x}$ to the LiOX$\cdot$BF$_{3}$ additive (when
present) was 10:1. 
\label{HXXX}
}
\newcolumntype{C}{>{\centering\arraybackslash}X}
\begin{tabularx}{\textwidth}{lCCCCCC}
\toprule
\textbf{Group}    & \multicolumn{2}{c}{\textbf{Ultrasonication of CF$_{x}$}} & \textbf{Thickness} & \textbf{Binder} & \multicolumn{2}{c}{\textbf{Electrolyte}} \\
\textbf{ID} & \textbf{Solvent} & \textbf{Additive} & (${\mu}$m) &                  & \textbf{Salt} & \textbf{Solvent} \\
\midrule
H293  & EtOH & - & 100 & PAN   & LiBF$_{4}$  & PC:DME:DOL \\
H314  & EtOH & - & 250 & PAN   & LiBF$_{4}$  & PC:DME:DOL \\
H214  & EtOH & - & 100 & PVDF  & LiBF$_{4}$  & PC:DME:DOL \\
H321  & EtOH & - & 250 & PVDF  & LiBF$_{4}$  & PC:DME:DOL \\
H366  & EtOH & - & 100 & PAN   & LiPF$_{6}$  & PC:DME:DOL \\
H371  & EtOH & - & 100 & PAN   & LiClO$_{4}$ & PC:DME:DOL \\
H429  & EtOH & - & 100 & PAN   & LiBOB       & PC:DME:DOL \\
H101  & F$_{3}$B$\cdot$OEt$_{2}$   & - & 100 & PVDF     & LiBF$_{4}$ & PC:DME:DOL \\
H274  & F$_{3}$B$\cdot$OEt$_{2}$   & LiOX$\cdot$BF$_{3}$ & 100 & PAN     & LiBF$_{4}$ & PC:DME:DOL \\
H304  & \multicolumn{2}{c}{\textbf{no ultrasonication}} & 100 & PAN      & LiBF$_{4}$ & PC:DME:DOL \\
H179  & F$_{3}$B$\cdot$OEt$_{2}$  & LiOX$\cdot$BF$_{3}$ & 100 & PAN       & LiBF$_{4}$ &  DMSO:DOL  \\
H387  & EtOH & - & 100 & PVDF & LiBF$_{4}$ & EC:DMC \\
\bottomrule
\end{tabularx}
\end{table}

The composition of the cells in Table \ref{HXXX} reflects the time
evolution of our research toward increasingly high power Li-CF$_{x}$
batteries and toward understanding the key factors that control
power density. The higher the number in the cell group ID the later the
composition was explored.

\begin{figure}[H]
\includegraphics[width=10.5 cm]{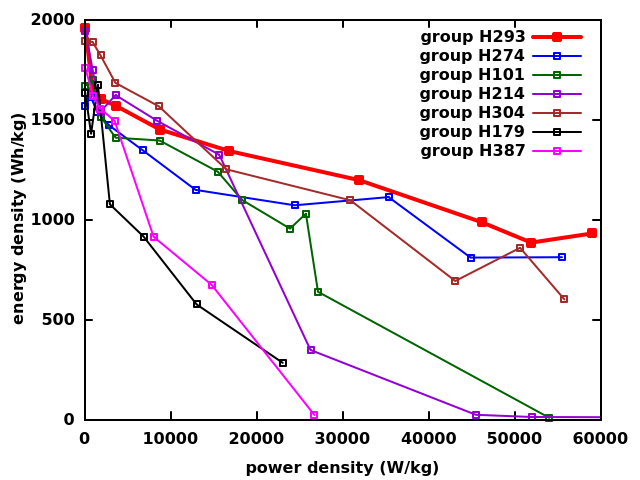}
\caption{
Gravimetric energy density vs power density of cells containing
cathodes with 100 ${\mu}m$ layer thickness. Each cell contained
a 1M LiBF$_{4}$ electrolyte. The
binder was either PAN or PVDF. The detailed 
composition of the cells can be found in Table \ref{HXXX}.
\label{EnergyVSPower}
}
\end{figure}

The gravimetric energy vs power density curves of cells with 100
${\mu}m$ thick cathodes and 1M LiBF$_{4}$ electrolytes are 
shown in Fig \ref{EnergyVSPower}. The values are given with
respect to the weight of the CF$_{x}$ in the cathode. 
These cells differ in the type of the binder (PAN or PVDF) 
and the preparation of the CF$_{x}$ 
active material (ultrasonication duration, solvent and additive).
The C rates of the discharge were between 0.05 and 40.

For a reference Li-CF$_{x}$ cell of traditional composition, 
we choose 
the same cell composition as the one in Ref \cite{li2021gaseous15C}
that investigated the effects of a gaseous BF$_{3}$ 
electrolyte additive.
It uses a PVDF binder and a 1 M LiBF$_{4}$ electrolyte in
EC:DMC(1:1). This configuration (group H387) 
is highly inferior in performance to all but
one of the cell configurations investigated here.

The best performing cells are from group H293. These cells
provide the highest energy density at very high power density
and perform consistently better than other cells at higher than 10
kW/kg power density. Group H293 cells have a simple composition:
their cathodes are composed of CF$_{x}$ that was ultrasonicated
in EtOH for 4 h and the binder was PAN, no additive or special
solvent (such as F$_{3}$B$\cdot$OEt$_{2}$) was applied.

\begin{figure}[H]
\includegraphics[width=10.5 cm]{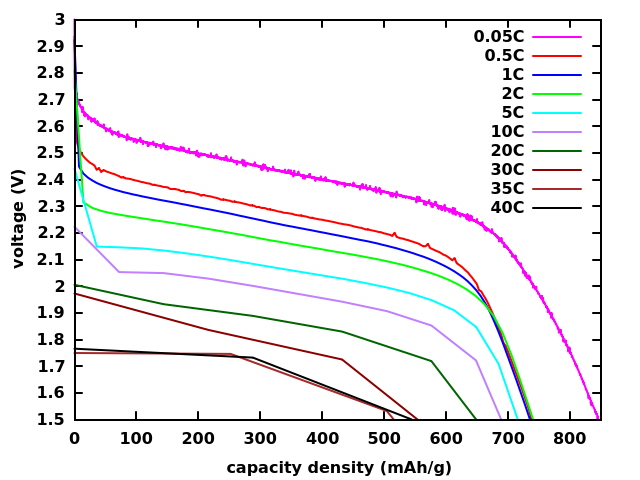}
\caption{
Voltage vs capacity density curves at various C rates for CF$_{x}$ 
batteries in our best performing cells that provide the highest
energy density at very high power densities (group H293). 
\label{PANVvsCap}
}
\end{figure}

The voltage vs capacity density curves during the discharge of
group H293 cells can be seen in Fig \ref{PANVvsCap}. They
compare very well (especially for high C rates) 
with similar curves in the best performing
CF$_{x}$ cells of the literature, 
such as those in Refs \cite{li2021gaseous15C,peng2021fluorinated20C,jiang2021electrochemical20C,wangfluorination20C,dai2014surface30C}
and especially those in Ref \cite{luo2021ultrafast50C} which we
consider the probably best performing CF$_{x}$ cell demonstrated to
date. While the CF$_{x}$ used in Ref \cite{luo2021ultrafast50C} 
is based on fluorinated graphene microspheres, 
our method uses the traditional and much easier to synthesize 
graphite fluoride. Other components of our cathode, such as the
PAN binder and the electrolyte (1 M LiBF$_{4}$ in
PC:DME:DOL(1:1:1)), are also conveniently available.
Therefore, our method appears to offer a more economic composition and
implementational simplicity while providing a competitive
performance.

Surprisingly, even the as purchased CF$_{x}$ (with no
ultrasonication, group H304) allows for relatively high energy
densities at high power densities when used with a PAN binder and
LiBF$_{4}$ electrolyte. However, when the PAN binder is exchanged to
PVDF, the the energy density rapidly decays at high power
densities (group H214). If the CF$_{x}$ is ultrasonicated in
F$_{3}$B$\cdot$OEt$_{2}$ then the performance becomes better
even if PVDF binder is used (group H101), however, it is still
much inferior to group H293. This indicates the beneficial
effects of BF$_{3}$ on the power density similarly to Ref
\cite{li2021gaseous15C}. 
The ultrasonication of CF$_{x}$ in F$_{3}$B$\cdot$OEt$_{2}$ with
the LiOX$\cdot$OEt$_{2}$ additive (group H274) does not improve the
performance of a simple ultrasonicated CF$_{x}$ active material
when PAN binder is used. This stresses again the robustness of the 
group H293 cathodes.

We have also explored a 1:1 volumetric mixture of DMSO and DOL
as a solvent in a 1M LiBF$_{4}$ electrolyte as it was found in earlier
literature that such an electrolyte can raise the discharge
voltage \cite{pang2015novel} at slow discharge. In our experience, 
this electrolyte has a poor performance at high C rates (high
power density) even when used with a PAN binder (group H179).

\begin{figure}[H]
\includegraphics[width=10.5 cm]{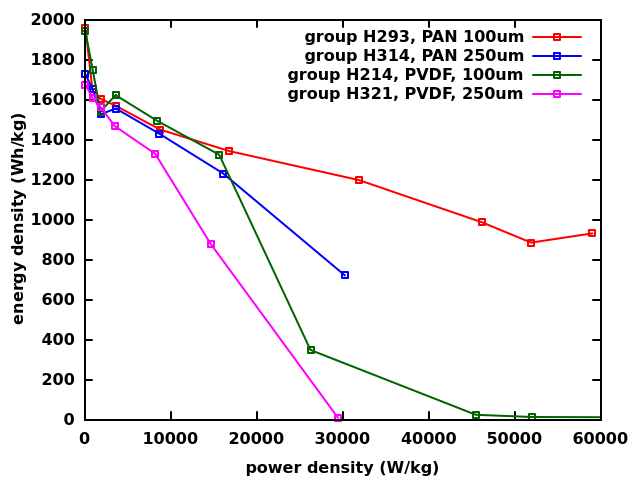}
\caption{
The dependence of the energy vs power density
performance on the type of the binder (PAN or PVDF) 
and on the thickness of the cathode (100 or 250 ${\mu}m$).
The electrolyte was 1 M LiBF$_{4}$ in PC:DME:DOL(1:1:1).
\label{PANvsPVDFvs100vs250um}
}
\end{figure}

Next, we have investigated the effect of cathode thickness on the
performance of Li-CF$_{x}$ cells while using either PAN or PVDF
binder and keeping the 1 M LiBF$_{4}$ electrolyte in
PC:DME:DOL(1:1:1). Two different cathode thickness were used,
100 ${\mu}m$ (1.0-1.5 mg/cm$^{2}$ CF$_{x}$ loading) and 250 ${\mu}m$
(3.0-4.6 mg/cm$^{2}$ CF$_{x}$ loading). Cathodes with PAN binder
have a greatly superior performance to cathodes with PVDF binder
at higher power densities (over 10 kW/kg), independent from
cathode thickness as shown in Fig \ref{PANvsPVDFvs100vs250um}. 

\begin{figure}[H]
\includegraphics[width=10.5 cm]{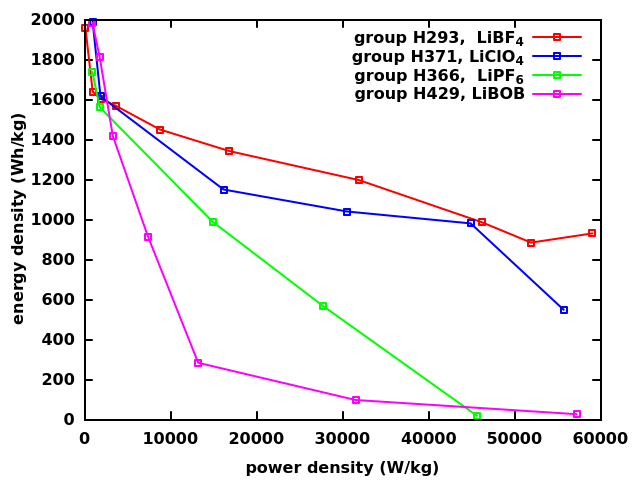}
\caption{
The dependence of the energy vs power density performance on
the type of the electrolyte. The effects of LiBF$_{4}$, LiClO$_{4}$, 
LiPF$_{6}$ and LiBOB electrolyte salts in PC:DME:DOL(1:1:1)
solvent are investigated. The cathode thickness was approximately 
100${\mu}m$ in all cases.
\label{Performance-vs-Electrolytes}
}
\end{figure}

We have also explored how the choice of the electrolyte salt
influences the performance of Li-CF$_{x}$ cells with PAN binder.
1 M solutions of LiBF$_{4}$, LiClO$_{4}$, LiPF$_{6}$ and LiBOB
were used in PC:DME:DOL(1:1:1). Fig \ref{Performance-vs-Electrolytes}
shows the dependence of the energy vs power density performance
on the type of the electrolyte. There is a divergent performance
at greater than 5 kW/kg power density: LiBF$_{4}$ performs best, followed
relatively closely by LiClO$_{4}$ while LiPF$_{6}$ performs
much less well and LiBOB is by far the slowest discharging one.

\begin{figure}[H]
\includegraphics[width=10.5 cm]{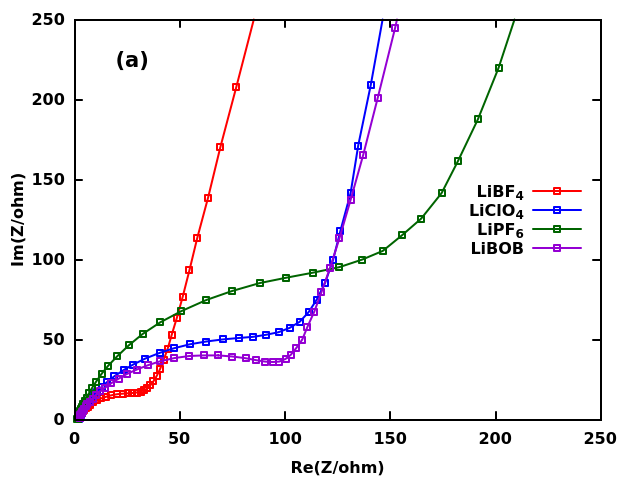}

\includegraphics[width=10.5 cm]{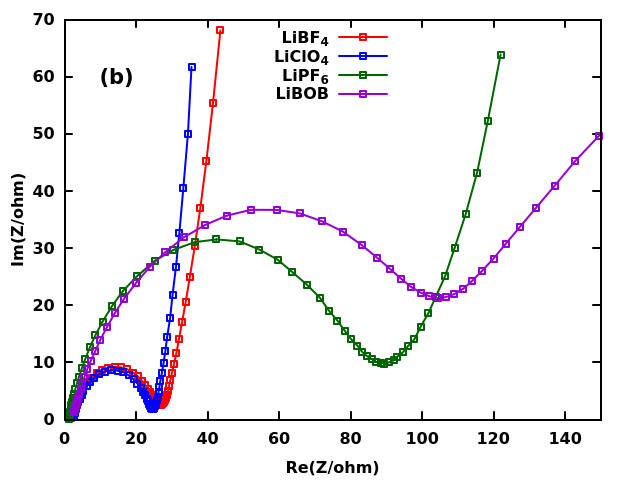}
\caption{
Impedance (Z) of Li-CF$_{x}$ cells with
1 M LiBF$_{4}$, LiClO$_{4}$, LiPF$_{6}$ and LiBOB
electrolytes before (panel \textbf{a}) and 1 h
after (panel \textbf{b}) full discharge at 0.5 C rate. 
The electrolyte solvent was PC:DME:DOL(1:1:1).
\label{EISElectrolytes}
}
\end{figure}

Figure \ref{EISElectrolytes}
shows the impedances (Z) of the electrolytes before 
(panel \textbf{a}) and 
1 h after (panel \textbf{b}) full discharge at 0.5 C rate.
The higher frequency impedances are close to the origin while the
low frequency ones are farther away.
The slope of the near straight section is
proportional to the diffusion coefficient of Li$^{+}$ ions and
the steeper the slope the higher the ionic conductivity of the
electrolyte. The maximum extent of the semicircle (formed after discharge)
on the Re(Z) axis
is related to the charge transfer resistance in the system: the
smaller the semi-circle the faster the charge transfer and the
higher the power density. The intercept of the impedance curve
with the Re(Z) axis is the bulk resistivity in the system (not
discussed here). The charge transfer is fastest in the LiBF$_{4}$
and LiClO$_{4}$ electrolytes while it is much slower in the 
LiPF$_{6}$ and LiBOB electrolytes. These observations are in
agreement with the measured power densities discussed above.

\begin{figure}[H]
\includegraphics[width=10.5 cm]{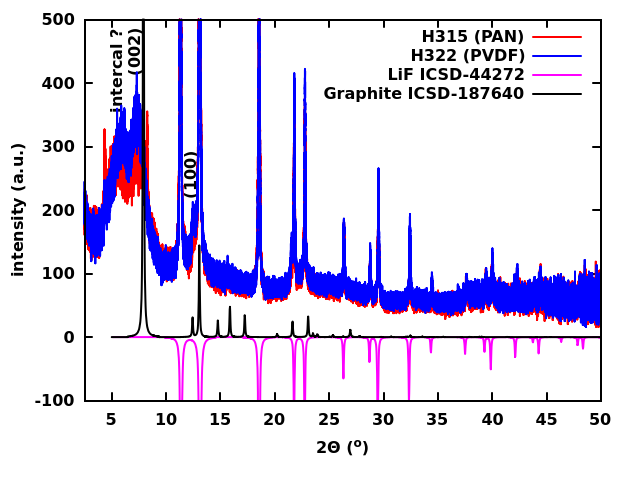}
\caption{
Synchrotron XRD patterns of the discharged cathodes of Li-CF$_{x}$ cells
with PAN and PVDF binders as compared to the patterns of
graphite and LiF. The intensities of the
LiF pattern are represented on the negative scale for clarity.
The approximate locations of the graphite (002) and (100)
reflections above the broad peaks are also indicated. These
broad peaks are associated with turbostratic graphite and
intercalated graphite.
\label{SXRD}
}
\end{figure}

Synchrotron XRD measurements were carried out on the discharged 
Li-CF$_{x}$ cells with PAN and PVDF binders in order to
investigate the extent of the hypothetical inhibition of LiF
crystallization by the PAN binder. 
The cells were discharged at 0.5 C rate using
250 ${\mu}m$ thick cathodes.
We found a large amount of crystalline LiF discharge
product, as indicated in Fig \ref{SXRD}. The broad peaks at
5.9-7.4 and 10-15 degrees region 
indicate the presence of intercalated graphite and 
turbostratic graphite like structures, respectively. 
Our experience is contradictory to that of 
Ref \cite{li2021gaseous15C} which 
found no crystalline LiF in the XRD patterns of the discharged
CF$_{x}$ cathodes when a small
concentration of BF$_{3}$ gas in the electrolyte was used.      
Therefore, we cannot confirm a similar degree of inhibition of
LiF formation when the PAN binder is used. It is however
possible that the inhibition caused by the PAN binder only slows
down the LiF formation on a shorter time scale during discharge
and therefore it could not be detected by our synchrotron XRD
measurements about three weeks after the discharge.
Consequently, the inhibition mechanism proposed by Jones and
Hossain in Ref \cite{jones2011polymer} may still be valid.

Also note that Jones and Hossain did not mention any
electrolyte effects on their proposed inhibition mechanism of
the LiF crystallization. Our
study points out the first time in the literature that a
combined effect of the binder and the electrolyte can be a
simple and robust means of greatly increasing the power density
of Li-CF$_{x}$ cells.

Further analysis of the XRD of the discharge products also
suggests that it may contain both turbostratic graphite (the
broad peak at 7.4 degrees near the (002) reflection of graphite) 
and a first stage intercalation complex of graphite 
(the broad peak at 5.9 degrees). 
Unfortunately, we could not further investigate 
if these broad peaks might 
be associated with hard carbon as suggested in 
Ref \cite{sayahpour2022revisiting}.

As the improvement of the power density is a combined effect of
the binder and the electrolyte, the understanding of the
interaction of PAN with
various electrolytes attracts our attention. It has been known for
a few decades that solid polymer electrolytes can be made from PAN and
LiClO$_{4}$ solution in some organic solvents \cite{watanabe1983ionic,chen2002polyacrylonitrile}.
The ionic conductivity of these electrolytes depend on the ratio
of their components and can be as high as
10$^{-3}$-10$^{-2}$ S/cm \cite{chen2002polyacrylonitrile}.
Potentially, the presence of such a polymer electrolyte in
our CF$_{x}$ cathodes may contribute to the 
enhanced discharge power density.

The mechanistic origin of the observed combined binder and electrolyte
effects on the power density is not clear yet. Several analogous
reactions however point toward the electro-catalytic effect of PAN in
the given circumstances. As PAN ((CH$_{2}$CH-CN)$_{x}$) 
has cyano (-CN) group side chains, it is
reasonable to assume that these -CN groups would be oxidized by CF$_{x}$
in a similar manner to the oxidation of NaCN by CF$_{x}$ \cite{siedle2022cyanographite}:
\begin{equation} \label{NCCN}
x NaCN + CF_{x} \rightarrow x NaF + C + x {\cdot}CN .
\end{equation}
The resulting ${\cdot}$CN radicals mostly dimerize to cyanogen (NC-CN) and
then polymerize to paracyanogen (CN)$_{2n}$
\cite{siedle2022cyanographite} and to a smaller extent may covalently
functionalize the graphene sheets \cite{bakandritsos2017cyanographene}.
In PAN, an oxidative effect by CF$_{x}$ is expected to result   
in the polymerization of the -CN groups in the side chain.   
Such a side chain polymerization of PAN has been known for long as an
effect of heating and it results in pyridine type rings and a ladder
type polymer \cite{chung1984optical}.

\begin{figure}[H]
\includegraphics[width=10.5 cm]{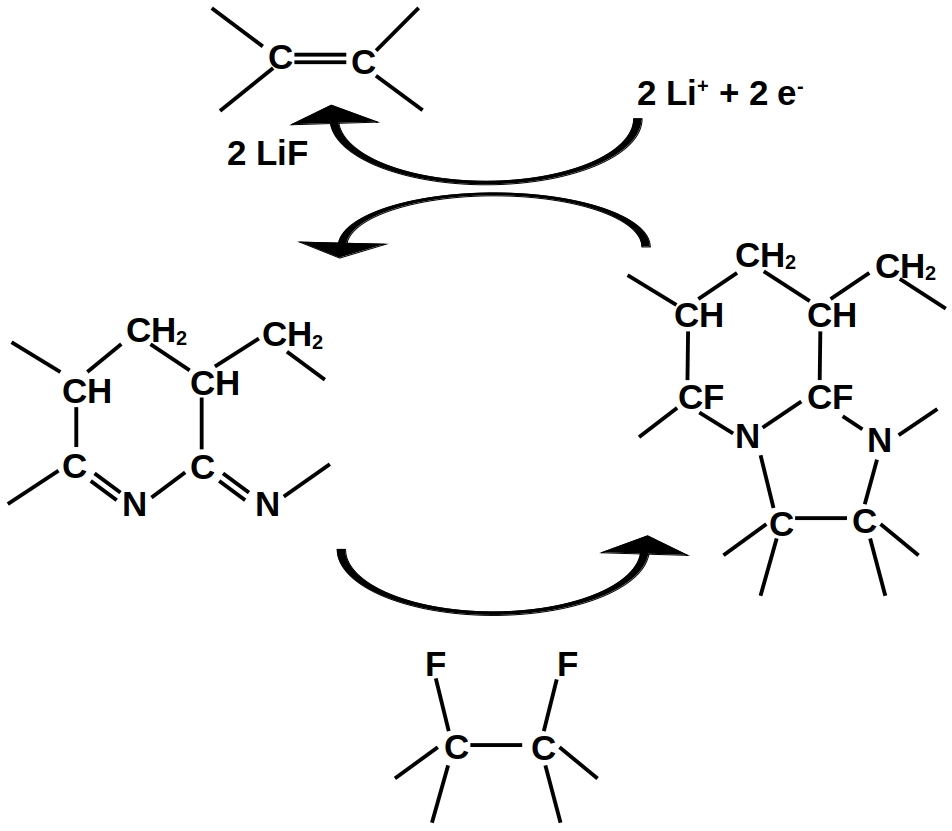}
\caption{
The proposed electro-catalytic action of the PAN binder on
CF$_{x}$ during the discharge of the Li-CF$_{x}$ cell.
PAN is represented in its side-chain polymerized form (PPAN), two
repeating units are shown with C=N double bonds. 
The N atoms of the PAN commit a nucleophilic attack on the C
atoms of CF$_{x}$. In the resulting activated complex, PAN is
bound to the CF$_{x}$ surface via C-N bonds while the F atoms
migrate to PAN and form new C-F bonds on PAN. The formation of
the activated complex is a cycloaddition 
involving C=N double bonds. When the battery
discharges, LiF forms, while the activated complex splits
into a residual CF$_{x}$ with a newly formed C=C double bond 
and the original PAN is recycled.
The cycle
can be repeated as long as there is sufficient Li and C-F supply.
\label{CatMech}
}
\end{figure}

We propose that the side chain polymerized PAN (PPAN) 
functions as an electro-catalyst. The catalytic mechanism is 
depicted in Fig \ref{CatMech}. PPAN and CF$_{x}$ form a
cyclo-adduct after the nucleophilic attack of PPAN on CF$_{x}$.
This cyclo-adduct is the activated complex of the reaction
mechanism. When the battery discharges, the cyclo-adduct is
reduced and splits into a residual CF$_{x}$ with a newly formed
C=C double bond and a recycled PPAN, while also LiF is formed.
The catalytic cycle can be active as long as there is a supply
of Li and C-F bonds. It seems, that PPAN must move on the
surface of CF$_{x}$ in order to harvest new C-F bonds. This
motion of PPAN can be beneficial for opening up the space between
stacked CF$_{x}$ layers. This may be the reason that the
catalytic effect works quite well even on non-ultrasonicated CF$_{x}$.
It is further expected that some electrolyte anions, such as 
PF$_{6}^{-}$ and BOB$^{-}$ may irreversibly 
react with PAN or PPAN and poison the catalyst. 
PF$_{6}^{-}$ may form P-N bonds with PPAN while 
the C=O double bonds in BOB$^{-}$ may form cyclo-adducts
with PAN. 
Other electrolytes, such as BF$_{4}^{-}$ and ClO$_{4}^{-}$
do not react irreversibly with PPAN and thus they 
do not poison the catalyst.
Further note that the pyridine-like units
in PPAN are also reminiscent of
the pyridine derivative 4-Dimethylaminopyridine (DMAP) which was 
successfully applied as a catalyst in the reaction of NaOH and
CF$_{x}$
and yielded -OH functionalized graphene (and presumably also
HO-OH, 
analogously to NC-CN of Eq \ref{NCCN}) \cite{bai2022unusual}.
As PVDF and PTFE binders are not
reactive with CF$_{x}$, they cannot provide catalytic effects.
Therefore, the proposed catalytic mechanism can account for all
the observed phenomena.

\section{Conclusions}
In the present study, we have demonstrated that the choice of the
binder and electrolyte plays a crucial role in achieving very
high power densities in Li-CF$_{x}$ batteries. As high as
931 Wh/kg energy density could be achieved at 59 kW/kg power
density in a coin cell 
when a PAN binder and LiBF$_{4}$ electrolyte are used.
While a former theoretical proposal by Jones and Hossain
predicted tremendous binder effects on the performance of
Li-CF$_{x}$ cells \cite{jones2011polymer} assuming that certain
binders will inhibit the crystallization of the discharge
product LiF, their proposal was
not aware of the role of the electrolyte in such effects.
The origin of the combined binder and electrolyte effect is not
clear yet. Based on the analogy to the proven oxidation of NaCN by
CF$_{x}$, we propose that PAN plays the role of an electro-catalyst in
the discharge of CF$_{x}$ as long as an electrolyte is available that
is able to reversibly bind to PAN.
Our method represents a simple and efficient route to very high power 
primary Li-CF$_{x}$ batteries.

\vspace{6pt} 


\authorcontributions{Conceptualization and experiment planning: 
K.N.; experimental work: H.H and K.N.; analysis, K.N., H.H. and
L.S.; original draft preparation and proposed electro-catalytic
mechanism: K.N.; review and editing: K.N., H.H. and L.S.}

\funding{This research was funded in most part by the
U.S. National Science Foundation under a STTR Phase I grant to
Boron Nitride Power LLC and IIT (award number 2109286).
Additional private funding was provided by Boron Nitride Power
LLC.}

\dataavailability{Data generated in the present work is
available upon request from the authors.} 

\acknowledgments{
This research used resources (particularly synchrotron XRD facilities) 
of the Advanced Photon Source, a U.S.
Department of Energy (DOE) Office of Science user facility operated for
the DOE Office of Science by Argonne National Laboratory under Contract
No. DE-AC02-06CH11357.
}

\conflictsofinterest{The authors declare no conflict of interest. The
funders had no role in the design
of the study; in the collection, analyses, or interpretation of data; in
the writing of the manuscript, or
in the decision to publish the results.} 


\begin{adjustwidth}{-\extralength}{0cm}

\reftitle{References}




\begin{thebibliography}{999}

\bibitem[Trahey \em{et~al.}(2020)Trahey, Brushett, Balsara, Ceder, Cheng,
  Chiang, Hahn, Ingram, Minteer, Moore, et~al.]{trahey2020energy}
Trahey, L.; Brushett, F.R.; Balsara, N.P.; Ceder, G.; Cheng, L.; Chiang, Y.M.;
  Hahn, N.T.; Ingram, B.J.; Minteer, S.D.; Moore, J.S.;  et~al.
\newblock Energy storage emerging: A perspective from the Joint Center for
  Energy Storage Research.
\newblock {\em Proceedings of the National Academy of Sciences} {\bf 2020},
  {\em 117},~12550--12557.

\bibitem[Yang \em{et~al.}(2021)Yang, Liu, Ge, Rountree, and
  Wang]{yang2021challenges}
Yang, X.G.; Liu, T.; Ge, S.; Rountree, E.; Wang, C.Y.
\newblock Challenges and key requirements of batteries for electric vertical
  takeoff and landing aircraft.
\newblock {\em Joule} {\bf 2021}, {\em 5},~1644--1659.

\bibitem[Bills \em{et~al.}(2020)Bills, Sripad, Fredericks, Singh, and
  Viswanathan]{bills2020performance}
Bills, A.; Sripad, S.; Fredericks, W.L.; Singh, M.; Viswanathan, V.
\newblock Performance metrics required of next-generation batteries to
  electrify commercial aircraft.
\newblock {\em ACS Energy Letters} {\bf 2020}, {\em 5},~663--668.

\bibitem[Krause \em{et~al.}(2021)Krause, Ruiz, Jones, Brandon, Darcy, Iannello,
  and Bugga]{krause2021performance}
Krause, F.; Ruiz, J.; Jones, S.; Brandon, E.; Darcy, E.; Iannello, C.; Bugga,
  R.
\newblock Performance of commercial Li-ion cells for future NASA missions and
  aerospace applications.
\newblock {\em Journal of The Electrochemical Society} {\bf 2021}, {\em
  168},~040504.

\bibitem[Krause \em{et~al.}(2018)Krause, Jones, Jones, Pasalic, Billings, West,
  Smart, Bugga, Brandon, and Destephen]{krause2018high}
Krause, F.C.; Jones, J.P.; Jones, S.C.; Pasalic, J.; Billings, K.J.; West,
  W.C.; Smart, M.C.; Bugga, R.V.; Brandon, E.J.; Destephen, M.
\newblock High specific energy lithium primary batteries as power sources for
  deep space exploration.
\newblock {\em Journal of the Electrochemical Society} {\bf 2018}, {\em
  165},~A2312.

\bibitem[Ndzebet \em{et~al.}(2018)Ndzebet, Destephen, Zhang, and
  Darch]{ndzebethigh2018}
Ndzebet, E.; Destephen, M.; Zhang, D.; Darch, D.
\newblock High Power and High Rate Li/CFx-MnO2 Pouch Cell Hybrid Technology.
\newblock In Proceedings of the 48th Power Sources Conference; ,  2018; pp.
  558--561.

\bibitem[Greatbatch \em{et~al.}(1996)Greatbatch, Holmes, Takeuchi, and
  Ebel]{greatbatch1996lithium}
Greatbatch, W.; Holmes, C.; Takeuchi, E.; Ebel, S.
\newblock Lithium/carbon monofluoride (Li/CFx): a new pacemaker battery.
\newblock {\em Pacing and clinical electrophysiology} {\bf 1996}, {\em
  19},~1836--1840.

\bibitem[Ruff and Bretschneider(1934)]{ruff1934reaktionsprodukte}
Ruff, O.; Bretschneider, O.
\newblock Die Reaktionsprodukte der verschiedenen Kohlenstoffformen mit Fluor
  II (Kohlenstoff-monofluorid).
\newblock {\em Zeitschrift f{\"u}r anorganische und allgemeine Chemie} {\bf
  1934}, {\em 217},~1--18.

\bibitem[R{\"u}dorff and R{\"u}dorff(1947)]{rudorff1947konstitution}
R{\"u}dorff, W.; R{\"u}dorff, G.
\newblock Zur Konstitution des Kohlenstoff-Monofluorids.
\newblock {\em Zeitschrift f{\"u}r anorganische Chemie} {\bf 1947}, {\em
  253},~281--296.

\bibitem[Watanabe and Fukuda(1970)]{watanabe1970primary}
Watanabe, N.; Fukuda, M.
\newblock Primary cell for electric batteries.
\newblock {\em US Patent 3536532} {\bf 1970}.

\bibitem[Watanabe and Fukuda(1972)]{watanabe1972high}
Watanabe, K.; Fukuda, M.
\newblock High energy density battery.
\newblock {\em US Patent 3700502} {\bf 1972}.

\bibitem[Fukuda \em{et~al.}(1981)Fukuda, Iijima, and
  Toyoguchi]{fukuda1981active}
Fukuda, M.; Iijima, T.; Toyoguchi, Y.
\newblock Active material for positive electrode of battery.
\newblock {\em US Patent 4271242} {\bf 1981}.

\bibitem[Sharma \em{et~al.}(2021)Sharma, Dubois, Gu{\'e}rin, Pischedda, and
  Radescu]{sharma2021fluorinated}
Sharma, N.; Dubois, M.; Gu{\'e}rin, K.; Pischedda, V.; Radescu, S.
\newblock Fluorinated (Nano) Carbons: CFx Electrodes and CFx-Based Batteries.
\newblock {\em Energy Technology} {\bf 2021}, {\em 9},~2000605.

\bibitem[Ahmad \em{et~al.}(2021)Ahmad, Batisse, Chen, and
  Dubois]{ahmad2021preparation}
Ahmad, Y.; Batisse, N.; Chen, X.; Dubois, M.
\newblock Preparation and Applications of Fluorinated Graphenes.
\newblock {\em C} {\bf 2021}, {\em 7},~20.

\bibitem[Wang \em{et~al.}(2022)Wang, Wang, Zhang, Cui, Yu, and
  Shi]{wang2022composite}
Wang, D.; Wang, G.; Zhang, M.; Cui, Y.; Yu, J.; Shi, S.
\newblock Composite cathode materials for next-generation lithium fluorinated
  carbon primary batteries.
\newblock {\em Journal of Power Sources} {\bf 2022}, {\em 541},~231716.

\bibitem[Groult and Tressaud(2018)]{groult2018use}
Groult, H.; Tressaud, A.
\newblock Use of inorganic fluorinated materials in lithium batteries and in
  energy conversion systems.
\newblock {\em Chemical Communications} {\bf 2018}, {\em 54},~11375--11382.

\bibitem[Zhang \em{et~al.}(2015)Zhang, Takeuchi, Takeuchi, and
  Marschilok]{zhang2015progress}
Zhang, Q.; Takeuchi, K.J.; Takeuchi, E.S.; Marschilok, A.C.
\newblock Progress towards high-power Li/CF x batteries: electrode
  architectures using carbon nanotubes with CF x.
\newblock {\em Physical Chemistry Chemical Physics} {\bf 2015}, {\em
  17},~22504--22518.

\bibitem[Chen \em{et~al.}(2021)Chen, Jiang, Jiang, Zou, Ran, Wang, Niu, and
  Wang]{chen2021fluorinated}
Chen, P.; Jiang, C.; Jiang, J.; Zou, J.; Ran, Q.; Wang, X.; Niu, X.; Wang, L.
\newblock Fluorinated Carbons as Rechargeable Li-Ion Battery Cathodes in the
  Voltage Window of 0.5--4.8 V.
\newblock {\em ACS Applied Materials \& Interfaces} {\bf 2021}, {\em
  13},~30576--30582.

\bibitem[Liu \em{et~al.}(2014)Liu, Li, Xie, and Fu]{liu2014rechargeable}
Liu, W.; Li, H.; Xie, J.Y.; Fu, Z.W.
\newblock Rechargeable room-temperature CF x-sodium battery.
\newblock {\em ACS Applied Materials \& Interfaces} {\bf 2014}, {\em
  6},~2209--2212.

\bibitem[Whittingham(1975)]{whittingham1975mechanism}
Whittingham, M.S.
\newblock Mechanism of reduction of the fluorographite cathode.
\newblock {\em Journal of The Electrochemical Society} {\bf 1975}, {\em
  122},~526.

\bibitem[Watanabe \em{et~al.}(1987)Watanabe, Nakajima, and
  Hagiwara]{watanabe1987discharge}
Watanabe, N.; Nakajima, T.; Hagiwara, R.
\newblock Discharge reaction and overpotential of the graphite fluoride cathode
  in a nonaqueous lithium cell.
\newblock {\em Journal of Power Sources} {\bf 1987}, {\em 20},~87--92.

\bibitem[Watanabe \em{et~al.}(1982)Watanabe, Hagiwara, Nakajima, Touhara, and
  Ueno]{watanabe1982solvents}
Watanabe, N.; Hagiwara, R.; Nakajima, T.; Touhara, H.; Ueno, K.
\newblock Solvents effects on electrochemical characteristics of graphite
  fluoride—lithium batteries.
\newblock {\em Electrochimica Acta} {\bf 1982}, {\em 27},~1615--1619.

\bibitem[Jang \em{et~al.}(2011)Jang, Liu, Neff, Yu, Wang, Xiong, and
  Zhamu]{jang2011graphene}
Jang, B.Z.; Liu, C.; Neff, D.; Yu, Z.; Wang, M.C.; Xiong, W.; Zhamu, A.
\newblock Graphene surface-enabled lithium ion-exchanging cells:
  next-generation high-power energy storage devices.
\newblock {\em Nano Letters} {\bf 2011}, {\em 11},~3785--3791.

\bibitem[Liu \em{et~al.}(2014)Liu, Zhamu, Neff, and Jang]{liu2014lithium}
Liu, C.; Zhamu, A.; Neff, D.; Jang, B.Z.
\newblock Lithium super-battery with a functionalized nano graphene cathode,
  2014.
\newblock US Patent 8,795,899.

\bibitem[Kim \em{et~al.}(2014{\natexlab{a}})Kim, Park, Hong, and
  Kang]{kim2014all}
Kim, H.; Park, K.Y.; Hong, J.; Kang, K.
\newblock All-graphene-battery: bridging the gap between supercapacitors and
  lithium ion batteries.
\newblock {\em Scientific reports} {\bf 2014}, {\em 4},~5278.

\bibitem[Kim \em{et~al.}(2014{\natexlab{b}})Kim, Park, Park, Lim, Hong, and
  Kang]{kim2014novel}
Kim, H.; Park, Y.U.; Park, K.Y.; Lim, H.D.; Hong, J.; Kang, K.
\newblock Novel transition-metal-free cathode for high energy and power sodium
  rechargeable batteries.
\newblock {\em Nano Energy} {\bf 2014}, {\em 4},~97--104.

\bibitem[Kornilov \em{et~al.}(2022)Kornilov, Penki, Cheglakov, and
  Aurbach]{kornilov2022li}
Kornilov, D.; Penki, T.R.; Cheglakov, A.; Aurbach, D.
\newblock Li/graphene oxide primary battery system and mechanism.
\newblock {\em Battery Energy} {\bf 2022}, {\em 1},~20210002.

\bibitem[Zhang \em{et~al.}(2009)Zhang, Foster, Wolfenstine, and
  Read]{zhang2009electrochemical}
Zhang, S.S.; Foster, D.; Wolfenstine, J.; Read, J.
\newblock Electrochemical characteristic and discharge mechanism of a primary
  Li/CFx cell.
\newblock {\em Journal of Power Sources} {\bf 2009}, {\em 187},~233--237.

\bibitem[Sayahpour \em{et~al.}(2022)Sayahpour, Hirsh, Bai, Schorr, Lambert,
  Mayer, Bao, Cheng, Zhang, Leung, et~al.]{sayahpour2022revisiting}
Sayahpour, B.; Hirsh, H.; Bai, S.; Schorr, N.B.; Lambert, T.N.; Mayer, M.; Bao,
  W.; Cheng, D.; Zhang, M.; Leung, K.;  et~al.
\newblock Revisiting discharge mechanism of CFx as a high energy density
  cathode material for lithium primary battery.
\newblock {\em Advanced Energy Materials} {\bf 2022}, {\em 12},~2103196.

\bibitem[Yazami \em{et~al.}(2007)Yazami, Hamwi, Gu{\'e}rin, Ozawa, Dubois,
  Giraudet, and Masin]{yazami2007fluorinated}
Yazami, R.; Hamwi, A.; Gu{\'e}rin, K.; Ozawa, Y.; Dubois, M.; Giraudet, J.;
  Masin, F.
\newblock Fluorinated carbon nanofibres for high energy and high power
  densities primary lithium batteries.
\newblock {\em Electrochemistry communications} {\bf 2007}, {\em
  9},~1850--1855.

\bibitem[Lam and Yazami(2006)]{lam2006physical}
Lam, P.; Yazami, R.
\newblock Physical characteristics and rate performance of (CFx) n (0.33< x<
  0.66) in lithium batteries.
\newblock {\em Journal of Power Sources} {\bf 2006}, {\em 153},~354--359.

\bibitem[Luo \em{et~al.}(2021)Luo, Wang, Chen, Chang, Xie, Ma, Lei, Pan, Pan,
  and Huang]{luo2021ultrafast50C}
Luo, Z.; Wang, X.; Chen, D.; Chang, Q.; Xie, S.; Ma, Z.; Lei, W.; Pan, J.; Pan,
  Y.; Huang, J.
\newblock Ultrafast Li/fluorinated graphene primary batteries with high energy
  density and power density.
\newblock {\em ACS Applied Materials \& Interfaces} {\bf 2021}, {\em
  13},~18809--18820.

\bibitem[Dai \em{et~al.}(2014)Dai, Cai, Wu, Yang, Xie, Wen, Zheng, and
  Zhu]{dai2014surface30C}
Dai, Y.; Cai, S.; Wu, L.; Yang, W.; Xie, J.; Wen, W.; Zheng, J.C.; Zhu, Y.
\newblock Surface modified CFx cathode material for ultrafast discharge and
  high energy density.
\newblock {\em Journal of Materials Chemistry A} {\bf 2014}, {\em
  2},~20896--20901.

\bibitem[Peng \em{et~al.}(2021)Peng, Kong, Li, Fu, Sun, Feng, and
  Feng]{peng2021fluorinated20C}
Peng, C.; Kong, L.; Li, Y.; Fu, H.; Sun, L.; Feng, Y.; Feng, W.
\newblock Fluorinated graphene nanoribbons from unzipped single-walled carbon
  nanotubes for ultrahigh energy density lithium-fluorinated carbon batteries.
\newblock {\em Science China Materials} {\bf 2021}, {\em 64},~1367--1377.

\bibitem[Jiang \em{et~al.}(2021)Jiang, Huang, Lu, and
  Liu]{jiang2021electrochemical20C}
Jiang, S.; Huang, P.; Lu, J.; Liu, Z.
\newblock The electrochemical performance of fluorinated ketjenblack as a
  cathode for lithium/fluorinated carbon batteries.
\newblock {\em RSC advances} {\bf 2021}, {\em 11},~25461--25470.

\bibitem[Wang \em{et~al.}()Wang, Feng, Kong, Peng, Hu, Li, Li, and
  Feng]{wangfluorination20C}
Wang, K.; Feng, Y.; Kong, L.; Peng, C.; Hu, Y.; Li, W.; Li, Y.; Feng, W.
\newblock The fluorination of boron-doped graphene for CFx cathode with
  ultrahigh energy density.
\newblock {\em Energy \& Environmental Materials}, p. e12437.

\bibitem[Li \em{et~al.}(2021)Li, Xue, Sun, Yu, Li, and Chen]{li2021gaseous15C}
Li, Q.; Xue, W.; Sun, X.; Yu, X.; Li, H.; Chen, L.
\newblock Gaseous electrolyte additive BF3 for high-power Li/CFx primary
  batteries.
\newblock {\em Energy Storage Materials} {\bf 2021}, {\em 38},~482--488.

\bibitem[Rangasamy \em{et~al.}(2014)Rangasamy, Li, Sahu, Dudney, and
  Liang]{rangasamy2014pushing}
Rangasamy, E.; Li, J.; Sahu, G.; Dudney, N.; Liang, C.
\newblock Pushing the theoretical limit of Li-CF x batteries: a tale of
  bifunctional electrolyte.
\newblock {\em Journal of the American Chemical Society} {\bf 2014}, {\em
  136},~6874--6877.

\bibitem[Jones and Hossain(2011)]{jones2011polymer}
Jones, S.C.; Hossain, S.
\newblock Polymer materials as binder for a CFx cathode,  2011.
\newblock {US} Patent App. 13/010,431.

\bibitem[N{\'e}meth(2014)]{nemeth2014materials}
N{\'e}meth, K.
\newblock Materials design by quantum-chemical and other
  theoretical/computational means: Applications to energy storage and
  photoemissive materials.
\newblock {\em International Journal of Quantum Chemistry} {\bf 2014}, {\em
  114},~1031--1035.

\bibitem[Zhang \em{et~al.}(2016)Zhang, N{\'e}meth, Bare{\~n}o, Dogan, Bloom,
  and Shaw]{zhang2016experimental}
Zhang, F.; N{\'e}meth, K.; Bare{\~n}o, J.; Dogan, F.; Bloom, I.D.; Shaw, L.L.
\newblock Experimental and theoretical investigations of functionalized boron
  nitride as electrode materials for Li-ion batteries.
\newblock {\em RSC Advances} {\bf 2016}, {\em 6},~27901--27914.

\bibitem[N{\'e}meth(2018)]{nemeth2018simultaneous}
N{\'e}meth, K.
\newblock Simultaneous oxygen and boron trifluoride functionalization of
  hexagonal boron nitride: a designer cathode material for energy storage.
\newblock {\em Theoretical Chemistry Accounts} {\bf 2018}, {\em 137},~157.

\bibitem[N{\'e}meth(2021)]{nemeth2021radical}
N{\'e}meth, K.
\newblock Radical anion functionalization of two-dimensional materials as a
  means of engineering simultaneously high electronic and ionic conductivity
  solids.
\newblock {\em Nanotechnology} {\bf 2021}, {\em 32},~245709.

\bibitem[Nemeth(2020)]{nemeth2020-US10693137}
Nemeth, K.
\newblock Functionalized boron nitride materials as electroactive species in
  electrochemical energy storage devices.
\newblock {\em US Patent 10,693,137} {\bf 2020}.

\bibitem[Nemeth(2022)]{nemeth2022radicalAnion-US11453596B2}
Nemeth, K.
\newblock Radical Anion Functionalization of Two-dimensional Materials.
\newblock {\em US Patent 11,453,596 B2} {\bf 2022}.

\bibitem[Marshall \em{et~al.}(2021)Marshall, Zhenova, Roberts, Petchey, Zhu,
  Dancer, McElroy, Kendrick, and Goodship]{marshall2021solubility}
Marshall, J.E.; Zhenova, A.; Roberts, S.; Petchey, T.; Zhu, P.; Dancer, C.E.;
  McElroy, C.R.; Kendrick, E.; Goodship, V.
\newblock On the solubility and stability of polyvinylidene fluoride.
\newblock {\em Polymers} {\bf 2021}, {\em 13},~1354.

\bibitem[Zor \em{et~al.}(2021)Zor, Suba\c{s}{\i}, Haciu, Somer, and
  Afyon]{zor2021guide}
Zor, C.; Suba\c{s}{\i}, Y.; Haciu, D.; Somer, M.; Afyon, S.
\newblock Guide to water free lithium bis (oxalate) borate (LiBOB).
\newblock {\em The Journal of Physical Chemistry C} {\bf 2021}, {\em
  125},~11310--11317.

\bibitem[Zhang \em{et~al.}(2013)Zhang, Ma, Zhu, Che, and Xiao]{zhang2013two}
Zhang, M.; Ma, Y.; Zhu, Y.; Che, J.; Xiao, Y.
\newblock Two-dimensional transparent hydrophobic coating based on liquid-phase
  exfoliated graphene fluoride.
\newblock {\em Carbon} {\bf 2013}, {\em 63},~149--156.

\bibitem[Zeng \em{et~al.}(2018)Zeng, Peng, Yu, Lang, Cao, and
  Zou]{zeng2018dynamic}
Zeng, X.; Peng, Y.; Yu, M.; Lang, H.; Cao, X.; Zou, K.
\newblock Dynamic sliding enhancement on the friction and adhesion of graphene,
  graphene oxide, and fluorinated graphene.
\newblock {\em ACS applied materials \& interfaces} {\bf 2018}, {\em
  10},~8214--8224.

\bibitem[Huo(2022)]{huo2022high}
Huo, H.
\newblock High Energy High Power Primary Lithium Batteries with Graphite
  Fluoride and Functionalized Boron Nitride Cathodes.
\newblock Master's thesis, Illinois Institute of Technology,  2022.

\bibitem[Tatagari(2021)]{tatagari2021functionalized}
Tatagari, V.R.
\newblock A Functionalized 2D Boron Nitride Electrode for Rechargeable
  Batteries.
\newblock Master's thesis, Illinois Institute of Technology,  2021.

\bibitem[Li \em{et~al.}(2022)Li, Liu, Chen, Wu, Zhou, Shen, and
  Zhou]{li2022multi}
Li, Y.Y.; Liu, C.; Chen, L.; Wu, X.Z.; Zhou, P.F.; Shen, X.Y.; Zhou, J.
\newblock Multi-layered fluorinated graphene cathode materials for lithium and
  sodium primary batteries.
\newblock {\em Rare Metals} {\bf 2022}, pp. 1--14.

\bibitem[Zhao \em{et~al.}(2020)Zhao, Liu, Zheng, Deng, Warren, Zhang, and
  Archer]{zhao2020designing}
Zhao, Q.; Liu, X.; Zheng, J.; Deng, Y.; Warren, A.; Zhang, Q.; Archer, L.
\newblock Designing electrolytes with polymerlike glass-forming properties and
  fast ion transport at low temperatures.
\newblock {\em Proceedings of the National Academy of Sciences} {\bf 2020},
  {\em 117},~26053--26060.

\bibitem[Pang \em{et~al.}(2015)Pang, Ding, Sun, Liu, Hao, Wang, Liu, and
  Xu]{pang2015novel}
Pang, C.; Ding, F.; Sun, W.; Liu, J.; Hao, M.; Wang, Y.; Liu, X.; Xu, Q.
\newblock A novel dimethyl sulfoxide/1, 3-dioxolane based electrolyte for
  lithium/carbon fluorides batteries with a high discharge voltage plateau.
\newblock {\em Electrochimica Acta} {\bf 2015}, {\em 174},~230--237.

\bibitem[Watanabe \em{et~al.}(1983)Watanabe, Kanba, Nagaoka, and
  Shinohara]{watanabe1983ionic}
Watanabe, M.; Kanba, M.; Nagaoka, K.; Shinohara, I.
\newblock Ionic conductivity of hybrid films composed of polyacrylonitrile,
  ethylene carbonate, and LiClO$_{4}$.
\newblock {\em Journal of Polymer Science: Polymer Physics Edition} {\bf 1983},
  {\em 21},~939--948.

\bibitem[Chen \em{et~al.}(2002)Chen, Lin, and Chen]{chen2002polyacrylonitrile}
Chen, H.; Lin, F.; Chen, C.
\newblock Polyacrylonitrile electrolytes 1. A novel high-conductivity composite
  polymer electrolyte based on PAN, LiClO$_{4}$ and a-Al$_{2}$O$_{3}$.
\newblock {\em Solid State Ionics} {\bf 2002}, {\em 150},~327--335.

\bibitem[Siedle \em{et~al.}(2022)Siedle, Losovyj, Stein, Pink, and
  Werner-Zwanziger]{siedle2022cyanographite}
Siedle, A.; Losovyj, Y.; Stein, B.D.; Pink, M.; Werner-Zwanziger, U.
\newblock Cyanographite.
\newblock {\em The Journal of Physical Chemistry C} {\bf 2022}, {\em
  126},~3001--3008.

\bibitem[Bakandritsos \em{et~al.}(2017)Bakandritsos, Pykal, B{\l}o{\'n}ski,
  Jakubec, Chronopoulos, Pol{\'a}kov{\'a}, Georgakilas, {\v{C}}{\'e}pe,
  Tomanec, Ranc, Bourlinos, Zbo\v{r}il, and
  Otyepka]{bakandritsos2017cyanographene}
Bakandritsos, A.; Pykal, M.; B{\l}o{\'n}ski, P.; Jakubec, P.; Chronopoulos,
  D.D.; Pol{\'a}kov{\'a}, K.; Georgakilas, V.; {\v{C}}{\'e}pe, K.; Tomanec, O.;
  Ranc, V.;  et~al.
\newblock Cyanographene and graphene acid: emerging derivatives enabling
  high-yield and selective functionalization of graphene.
\newblock {\em ACS nano} {\bf 2017}, {\em 11},~2982--2991.

\bibitem[Chung \em{et~al.}(1984)Chung, Schlesinger, Etemad, Macdiarmid, and
  Heeger]{chung1984optical}
Chung, T.C.; Schlesinger, Y.; Etemad, S.; Macdiarmid, A.; Heeger, A.
\newblock Optical studies of pyrolyzed polyacrylonitrile.
\newblock {\em Journal of Polymer Science: Polymer Physics Edition} {\bf 1984},
  {\em 22},~1239--1246.

\bibitem[Bai \em{et~al.}(2022)Bai, Xu, Liu, Dong, Zhang, Li, and
  Zhao]{bai2022unusual}
Bai, L.; Xu, Y.; Liu, A.; Dong, L.; Zhang, K.; Li, W.S.; Zhao, F.G.
\newblock Unusual graphite fluoride hydrolysis toward unconventional graphene
  oxide for high-performance supercapacitors and Li-ion batteries.
\newblock {\em Chemical Engineering Journal} {\bf 2022}, {\em 434},~134639.

\end{thebibliography}

\end{adjustwidth}
\end{document}